\documentclass[a4paper,11pt]{article}
\usepackage{jinstpub} 
\usepackage{lineno}

\pdfoutput=1
\usepackage[utf8]{inputenc}
\usepackage{float}
\usepackage{lmodern}

\usepackage{newtxtext,newtxmath}
\usepackage{graphicx}
\usepackage{url}
\usepackage{hyperref}
\usepackage{color}
\usepackage{enumitem}
\usepackage{subfigure}
\usepackage[dvipsnames]{xcolor}
\hypersetup{allcolors=[rgb]{0.0 0.0 0.6},linkcolor=[rgb]{0.75 0.05 0.05}}
\usepackage{bm}
\usepackage{epsfig}

\usepackage{slashed}
\usepackage{color}
\usepackage{accents}
\usepackage{url}
\usepackage[dvipsnames]{xcolor}
\usepackage{cancel}
\usepackage[normalem]{ulem} 
\usepackage[colorinlistoftodos]{todonotes}
\usepackage{multirow}
\usepackage{natbib}
\usepackage{dsfont}
\usepackage{bbold}
\usepackage{xspace}
\usepackage{siunitx}
\DeclareSIUnit\sq{\ensuremath{\Box}} 

\newcommand{\um}{$\mu$m\xspace}
\newcommand{\uA}{$\mu$A\xspace}

\newcommand{\exclude}[1]{}

\begin{document}


\author[a]{Cristi\'an Pe\~na,} 
\emailAdd{cmorgoth@fnal.gov}
\affiliation[a]{Fermi National Accelerator Laboratory, Batavia, IL 60510, U.S.A.}

\author[a,b]{Christina Wang,}
\affiliation[b]{California Institute of Technology, Pasadena, CA 91125, U.S.A.}

\author[a,b]{Si Xie,} 

\author[b]{Adolf Bornheim,} 

\author[c,d]{Mat\'ias Barr\'ia}
\author[c,d]{Claudio San Mart\'in,}
\affiliation[c]{Departamento de F\'isica y Astronom\'ia, Universidad T\'ecnica Federico Santa Mar\'ia, Valpara\'iso 2390123, Chile}
\affiliation[d]{Centro Cient\'ifico Tecnol\'ogico de Valpara\'iso-CCTVal, Universidad T\'ecnica Federico Santa Mar\'ia, Casilla 110-V, Valpara\'iso, Chile}

\author[c,d]{Valentina Vega,}

\author[a]{Artur Apresyan,} 

\author[e]{Emanuel Knehr,}
\affiliation[e]{NASA Jet Propulsion Laboratory, Pasadena, CA 91011, USA}

\author[e,f]{Boris Korzh,}
\author[e,g]{Jamie Luskin}
\affiliation[g]{University of Maryland, College Park, MD 20742, USA}
\author[b]{Lautaro Narv\'{a}ez,}

\author[b]{Sahil Patel,}

\author[e]{Matthew Shaw,}

\author[b]{Maria Spiropulu} 

\affiliation[f]{University of Geneva, 1205 Geneva, Switzerland}


\title{High Energy Particle Detection with Large Area Superconducting Microwire Array}


\abstract{

We present the first detailed study of an 8-channel $2\times2$~mm$^{2}$ WSi superconducting microwire single photon detector (SMSPD) array exposed to 120~GeV proton beam and 8~GeV electron and pion beam at the Fermilab Test Beam Facility.
The SMSPD detection efficiency was measured for the first time for protons, electrons, and pions, enabled by the use of a silicon tracking telescope that provided precise spatial resolution of 30~\um for 120~GeV protons and 130~\um for 8~GeV electrons and pions.
The result demonstrated consistent detection efficiency across pixels and at different bias currents.
Time resolution of 1.15~ns was measured for the first time for SMSPD with proton, electron, and pions, enabled by the use of an MCP-PMT which provided a ps-level reference time stamp.
The results presented is the first step towards developing SMSPD array systems optimized for high energy particle detection and identification for future accelerator-based experiments.

}

\maketitle
{
  \hypersetup{linkcolor=black}
  \tableofcontents
}


\section{Introduction}

Superconducting Nanowire Single Photon Detectors (SNSPDs) are a leading detector technology for single-photon detection with diverse applications in optical communications~\cite{Mao:18,PhysRevLett.124.070501}, quantum information science~\cite{PRXQuantum.1.020317,Takesue:15, Shibata:14, PhysRevA.90.043804,Najafi_2015,Weston:16}, and astronomy~\cite{PhysRevA.97.032329,PhysRevLett.123.070504, 10.1117/1.JATIS.7.1.011004}.
SNSPDs have been demonstrated to have ultra-low energy threshold of below 0.04~eV (or above 29$\mu$m)~\cite{Taylor:23}, low dark counts of $10^{-5}$~Hz~\cite{7752769,Shibata:14, Chiles:2021gxk}, and pico-second level time resolution~\cite{Korzh:2020, Mueller:24}, making them a potentially game-changing detector technology in high energy physics applications at accelerator-based experiments. 


However, until recently, SNSPDs have had limited use cases in high energy physics (HEP) experiments due to the small active area (100~$\mu \text{m}^2$) of traditional SNSPDs which are comprised of nanowires with widths of 100~nm.
Recent advances in the preparation of thin superconducting films have enabled the fabrication of large area (mm$^2$) single photon detectors with micrometer-width superconducting wires~\cite{10.1063/5.0150282, 10.1063/5.0044057, 10.1117/1.JATIS.7.1.011004}, which we refer to as superconducting microwire single-photon detectors (SMSPDs) in the rest of the paper.
This advancement in large active area makes SMSPD an ideal photosensor to detect single photons in dark matter detection experiments~\cite{BREAD:2021tpx, Chiles:2021gxk} and a potential innovative detector technology for future accelerator-based experiments.

The fundamental operating principles of SNSPDs and SMSPDs suggests that energy deposit from charged particles similarly induce hotspot formation and can be detected efficiently~\cite{Sclafani_2012,10.1063/1.4740074,SUZUKI20082001, Cristiano_2015, 10.1063/1.3506692,Lee:2023brm}.
However, previous studies have been limited to low energy (keV) electrons and ions and with the conventional small area, single-pixel SNSPDs.
To characterize SMSPDs for a future accelerator-based experiments, such as EIC~\cite{AbdulKhalek:2022hcn}, FCC-ee~\cite{FCC:2018evy}, FCC-hh~\cite{FCC:2018vvp}, or the muon collider~\cite{MuonCollider:2022glg}, we make use of the GeV-energy particles from the Fermilab Test Beam Facility (FTBF)~\cite{ftbf}.

In this paper, we report for the first time a full characterization of the detection efficiency and time resolution of an 8-pixel $2\times2$ mm$^2$ SMSPD array fabricated tungsten silicide (WSi) films under the exposure of 120~GeV protons and 8~GeV electrons and pions.
The SMSPD array under test is described in Section~\ref{sec:snspd}.
Section~\ref{sec:ftbf} describes the experimental setup at FTBF. 
The results of the full characterization of the SMSPD array are presented in Section~\ref{sec:results}.
Finally, the summary is presented in Section~\ref{sec:summary}.

\section{Superconducting microwire single photon detector array}
\label{sec:snspd}
\begin{figure}[ht]
	\centering
	\includegraphics[width=1.00\linewidth]{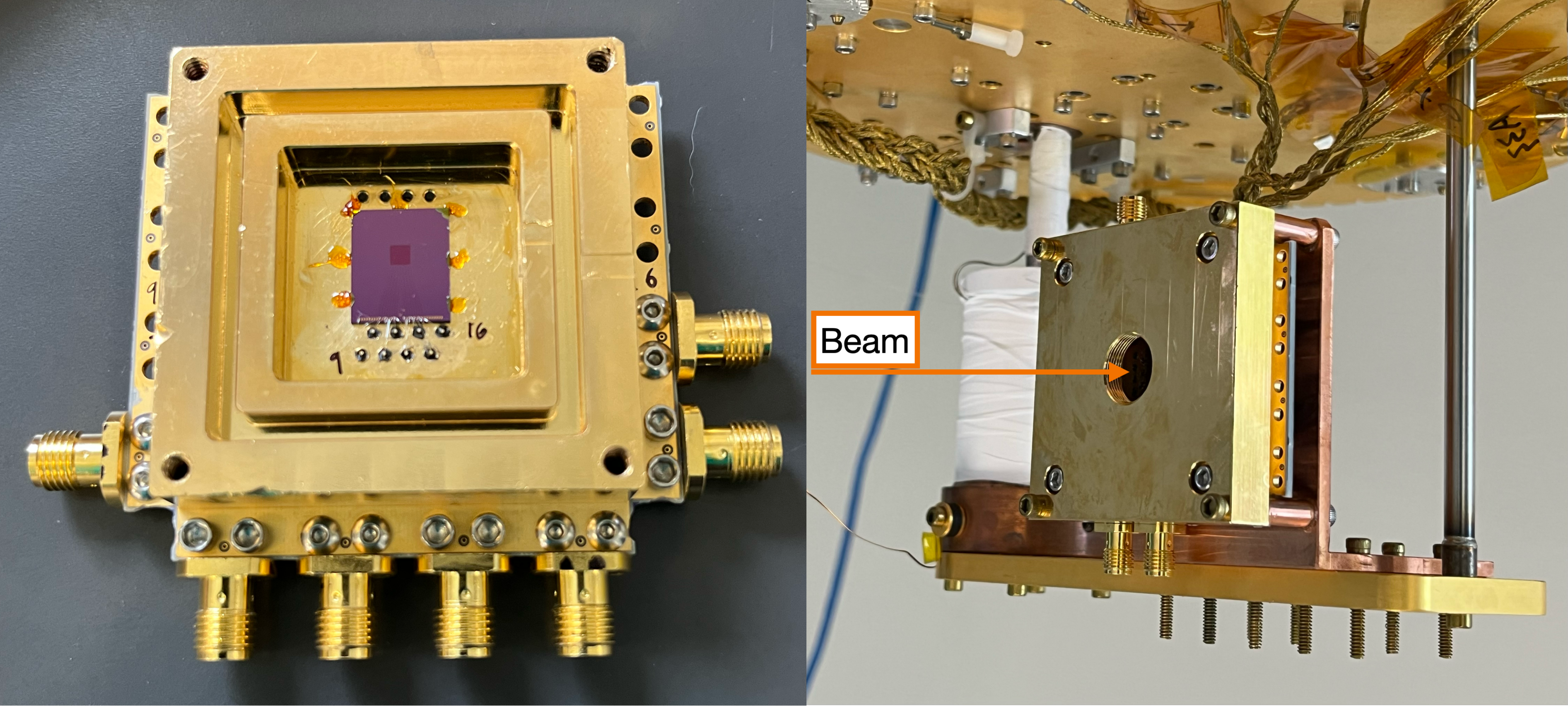}
	\includegraphics[width=0.95\linewidth]{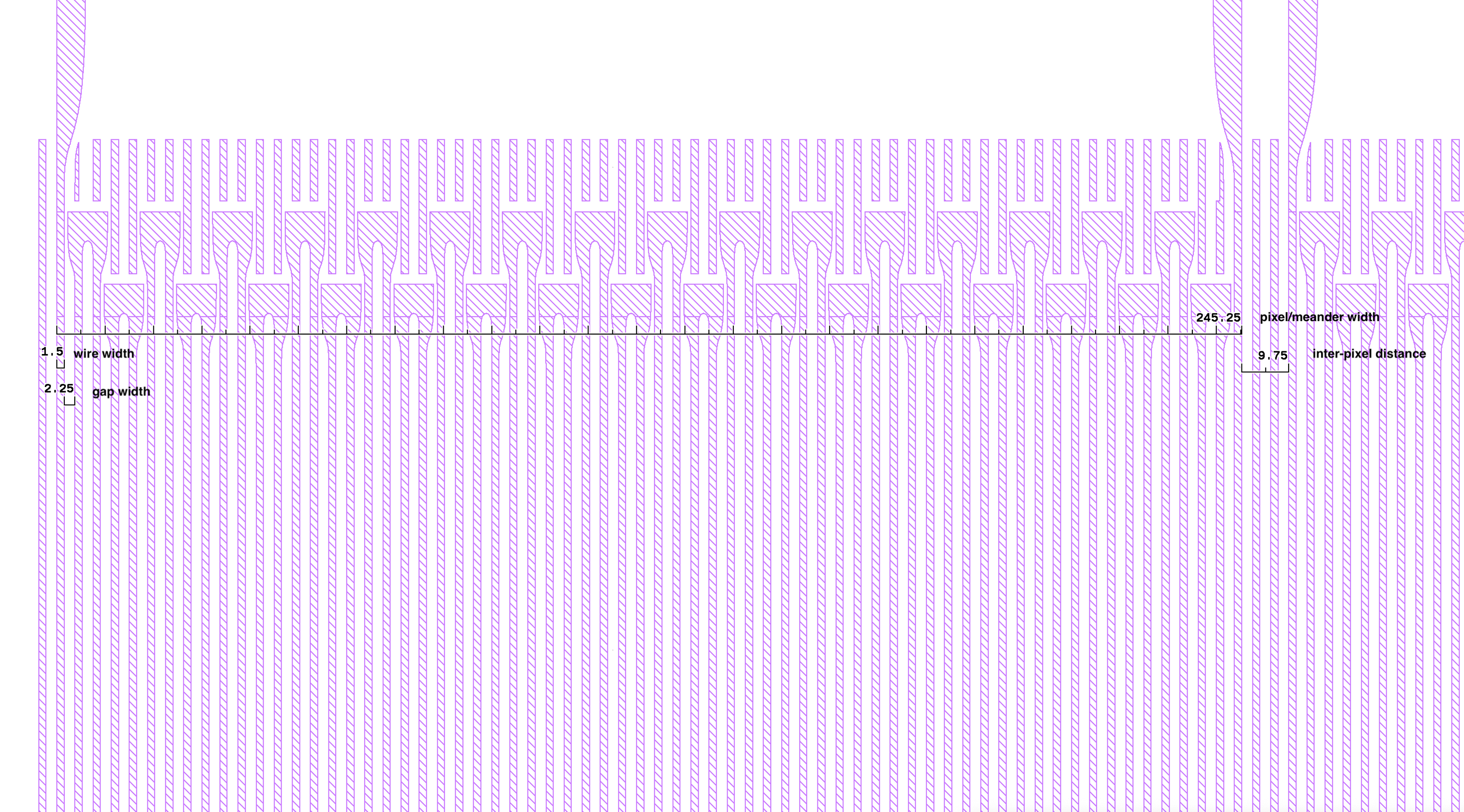}
 	\caption{Photographs of the SMSPD under study enclosed in a dark box attached to the cold plate in the cryostat (top left) and the SMSPD with the lid of the dark box opened (top right) and a schematic illustrating the meander structure of the sensor (bottom).
}
	\label{fig:Mounting}
\end{figure}

The detector under test is a $2\times2$~mm$^{2}$ 8-channel SMSPD array fabricated on 3~nm thick WSi film. The fabrication was carried out at the Jet Propulsion Laboratory.
Each pixel has a size of $0.25\times2$~mm$^{2}$.
The WSi film was sputtered from a $\text{W}_\text{50}\text{Si}_\text{50}$ target and deposited onto an oxidized silicon substrate with a 240~nm-thick oxide.
The sheet resistance is measured to be 1.2~\si{k\ohm}/\si{\sq} at room temperature.
The critical temperature is 2.8~K.
The thin WSi film was coated with 3~nm of amorphous silicon to prevent oxidation of the WSi film.
The SMSPD were patterned using optical lithography with 1.5~\um-wide wires meandering with a 2.25~\um gap width, amounting to a $40\%$ fill factor.
After etching, the microwires were covered with 40~nm of $\text{SiO}_\text{2}$ for passivation.
More details on the optical lithography of SMSPD can be found in Refs.~\cite{10.1063/5.0150282, 6994823}.
In this work, each pixel has an individual single-ended readout.
A photograph of the sensor, its meandering structure, and packaging enclosure is shown in Figure~\ref{fig:Mounting} (left).


The SMSPD is cooled down to 0.9~K with a PhotonSpot Helium Sorption Fridge that has a hold time of about 10-12 hours, allowing us to operate concurrently with the 12-hour beam time shifts allocated at FTBF.
The SMSPD was mounted vertically on a copper L-bracket from the base plate.
It was mechanically and thermally coupled to have normal incidence with respect to the beam, as shown in Figure~\ref{fig:Mounting} (right).

In this test beam campaign, we biased, amplified, and read out four neighboring pixels of the available eight pixels independently. 
A two-stage cryogenic DC-coupled amplifier operating at 40~K was developed by our group to provide a total gain of 30~dB for signal in the 100~MHz to 1~GHz frequency range, similar to that used in Ref.~\cite{10.1063/5.0150282}.
The first stage of the amplifier is based on a low noise high-electron-mobility transistor and the second stage is based on a silicon germanium amplifier.
The DC-coupled amplifier also simultaneously provides SMSPD biasing through the same signal cable.
The signal from each SMSPD pixel is connected through RF cables to room temperature, further amplified by a low noise, room temperature amplifier (Mini-Circuits ZFL-1000LN+), providing another 20~dB gain for signals in the 0.1~MHz to 1~GHz frequency range.
Subsequently, the RF signals are split into two, one of which is recorded by a high-rate time-to-digital converter (Swabian Time Tagger) and the other waveform recorded by an oscilloscope, as described in more detail in Section~\ref{sec:ftbf}. 

\begin{figure}[ht]
	\centering
	\includegraphics[width=0.6\linewidth]{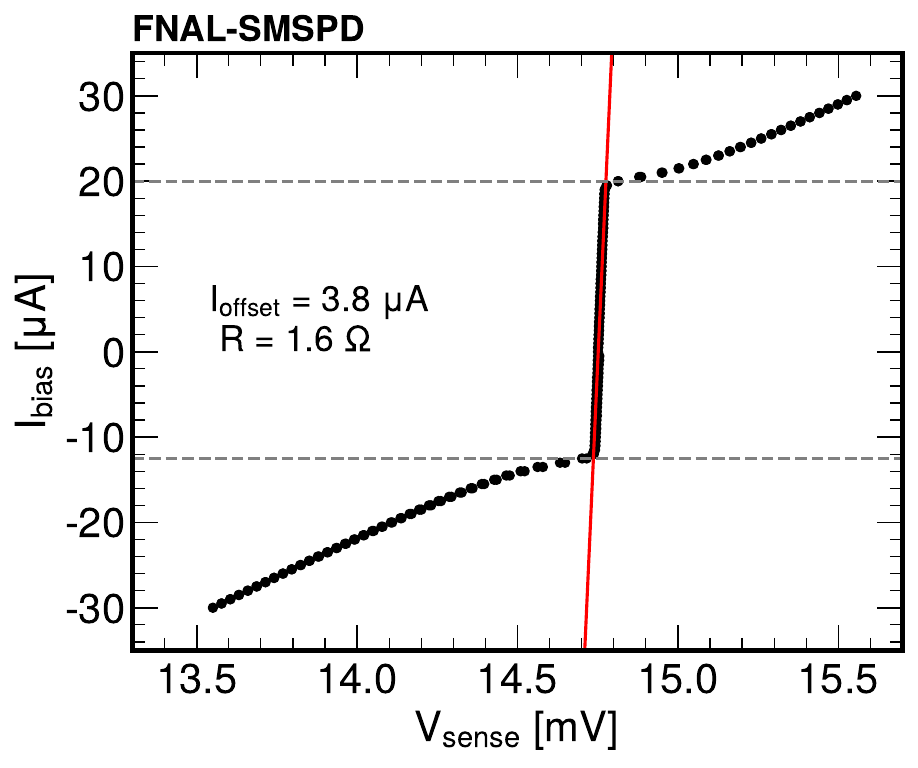}
 	\caption{The bias current ($I_\text{bias}$) with respect to voltage across one of the SMSPD pixels ($V_\text{sense}$). The linear region in the middle is the region when the SMSPD is superconducting. 
    }
	\label{fig:IVCurve}
\end{figure}

Since the bias current of the SMSPD is provided through the DC-coupled amplifier, it not only depends on the bias voltage provided, but also depends on the DC offset of the amplifier and the parasitic resistance of the SMSPD, both of which might vary between different cool downs.
Therefore, the current–voltage characteristic (IV curve) of the SMSPD pixels were measured at the beginning of the test beam campaign to derive the DC offset ($I_\text{offset}$) of the cryogenic amplifier and the parasitic resistance ($R$) of the SMSPD, as shown in Figure~\ref{fig:IVCurve}.
The IV curve for one of the SMSPD pixels was also measured at the beginning of every cool down to monitor variations in the parasitic resistance of the SMSPD and the DC offset of the cryogenic amplifier across cool downs, resulting in an uncertainty of 0.32~\uA in the SMSPD bias current.

Previous measurements~\cite{10.1063/5.0150282} have shown that similar SMSPD arrays reach saturated internal detection efficiency for photons with wavelength of 1064~nm. 
However, it is known that the time resolution of SMSPD arrays with wider and longer wires are not optimized, especially for single-ended readout.
For single-ended readout, the time delay introduced at different longitudinal position of particle detection becomes significant as the length of the microwire increases~\cite{Korzh:2020}.
Based on previous measurements of SMSPD arrays~\cite{10.1063/5.0150282} using slightly narrower (1~\um) and shorter microwires (about two times shorter), the time resolution achieved was 325~ps.
Therefore, we expect the time resolution of the SMSPD array under test in this paper to be on the order of hundreds of ps to a few ns.

\section{The experimental setup at the FNAL Test Beam Facility}
\label{sec:ftbf}

The data presented in this paper were collected at the FTBF, which provides a 120~GeV primary proton beam bunched at 53~MHz from the Fermilab Main Injector accelerator. 
The beam is resonantly extracted in a slow spill for each main injector cycle delivering a single 4.2~s long spill per minute.
The primary proton beam can also be targeted to a 30~cm aluminum that is 145~m upstream of the experimental hall to create secondary particle beams with energies 1--32~GeV, consisting of a mixture of pions, muons, and/or electrons.
In this paper, we will report on the results of the SMSPD response to 120~GeV proton beam and 8~GeV secondary electron and pion beams consisting of about 35\% of pions and 65\% of electrons, with both beams tuned to yield approximately 50,000 particles per spill.

The FTBF is equipped with a silicon tracking telescope to measure the position of each incident particle. 
The telescope consists of 12 strip modules with 60~\um pitch, in alternating orientation along the x- and y-axes, and four recently-updated pixel layers using the CMS Phase 2 pixel detectors~\cite{CERN-LHCC-2017-009} with a pitch of $25\times100 \mu$m$^{2}$~\cite{Madrid:2022rqw, KWAN2016162}.
Since each SMSPD pixel is only 250~\um wide, it is important to optimize the spatial resolution of the telescope to be able to give an accurate measurement of the detection efficiency of the SMSPD.
To optimize the position resolution measurement from the telescope, the SMSPD is placed in between the telescope with 6 strip modules upstream, and 4 pixels and 6 additional strip modules downstream, where the modules are placed as close to the cryostat as possible. 
During this data-taking period, the spatial resolution of the telescope at the SMSPD was measured to be 30~\um$\pm$ 1~\um for 120~GeV protons and 130~\um$\pm$20~\um for 8~GeV electrons and pions.

The spatial resolution of the telescope for protons was estimated by studying the smearing of the efficiency turn-on curve at the straight edge of two Low Gain Avalanche Detectors (LGADs) that are placed immediately upstream and downstream of the cryostat, respectively.
Assuming the response of the LGAD at the edge of the active area is a step function with a turn-on sharper than 1--2~\um~\cite{Heller_2022}, then any smearing observed near the turn-on can be attributed to the tracker resolution and extracted by a fit to an error function. 
The same method used on the edge of the SMSPD sensor also resulted in a measured resolution of 30~\um, indicating that the impact of potential fabrication or sensor response imperfections of the SMSPD is negligible. 
Therefore, the same measurement technique used on the edge of the SMSPD sensor can be reliably applied to the 8~GeV electron and pion beam particles, which resulted in a measured resolution of 130~\um.

The telescope data acquisition hardware is based on the CAPTAN (Compact And Programmable daTa Acquisition Node) system developed at Fermilab. 
The CAPTAN is a flexible and versatile data acquisition system designed to meet the readout and control demands of a variety of pixel and strip detectors for high energy physics applications~\cite{CAPTAN}.

A Photek 240 micro-channel plate (MCP-PMT) detector operated at -4~kV is located downstream of the telescope and the SMSPD and provides a precise reference timestamp with resolution below 10~ps~\cite{Ronzhin:2015idh}.

The SMSPD and MCP-PMT waveforms are acquired using a Lecroy Waverunner 8208HD oscilloscope. 
This oscilloscope features eight readout channels with a bandwidth of 2~GHz and a sampling rate of 10~GS/s per channel. 
Its deep memory is particularly well suited for the FTBF beam structure, allowing a burst of 30000 events to be acquired during each 4.2~s spill and written to disk during the longer inter-spill period.

The trigger signal to both the telescope and the oscilloscope originates in an independent device.
For a proton beam that has a beam-spot with 1--2~mm width, a plastic scintillator coupled to a photomultiplier tube is used.
For the secondary electron and pion beam that has a larger beam-spot of 6~mm, we used a customized 4~mm$\times$5~mm LYSO crystal coupled to a silicon photomultiplier to limit the triggered events to a small area around the SMSPD.

\begin{figure}[ht]
	\centering
	\includegraphics[width=0.95\linewidth]{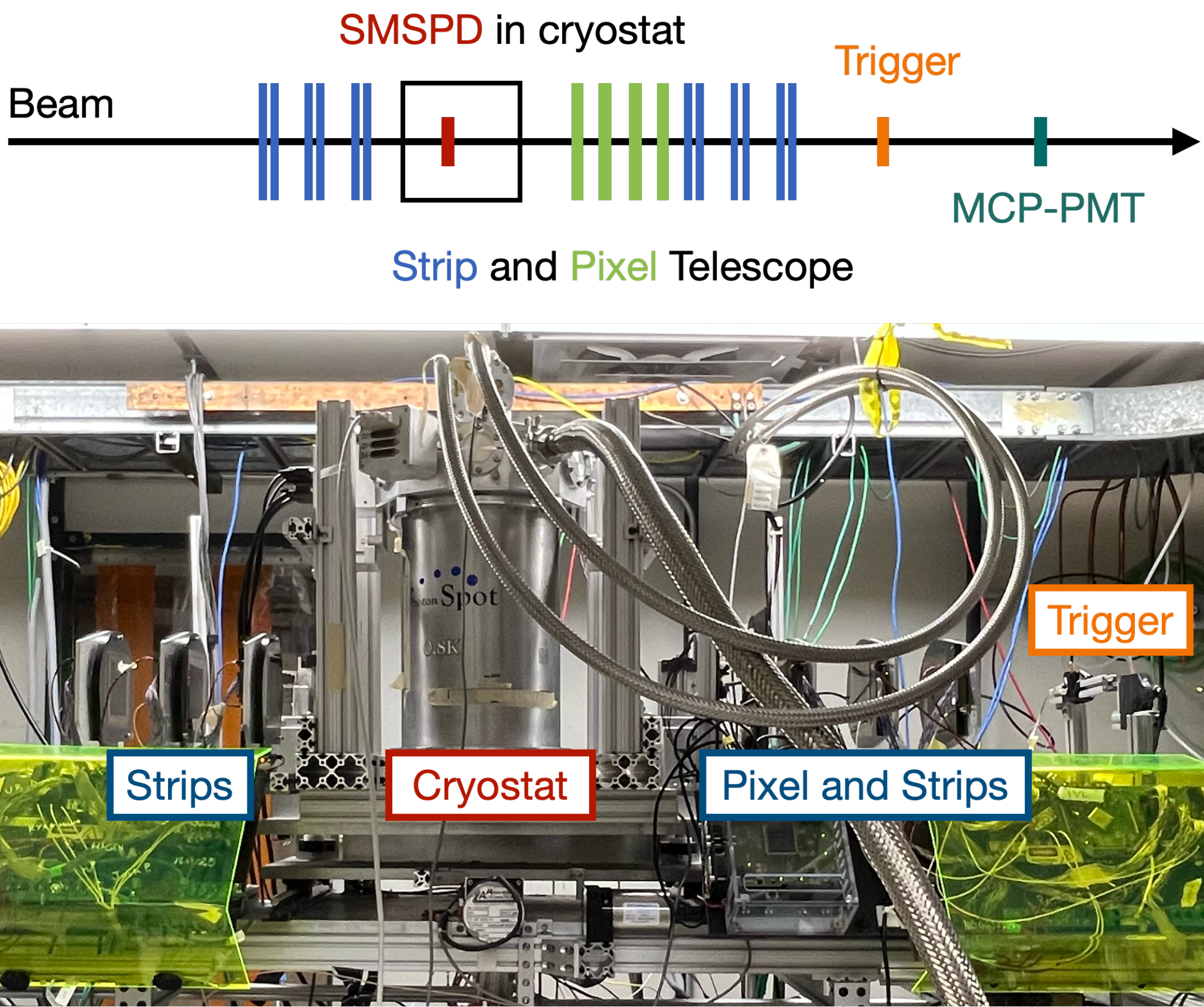}
 	\caption{
  A schematic diagram (top) and photograph (bottom) of the SMSPD under study and the reference instruments along the beamline. The MCP-PMT is further down the beamline and not shown in the photo.
}
	\label{fig:FTBFSetup}
 
\end{figure}

Events are built, merging the telescope and oscilloscope data offline, by matching trigger counters from each system.
A schematic of the experimental setup is shown on the top of Figure~\ref{fig:FTBFSetup}, which represents the arrangement of the cryostat hosting the SMSPD with respect to the telescope tracker, trigger, and MCP-PMT. 
The bottom of Figure~\ref{fig:FTBFSetup} shows a photograph of the setup.

\section{Experimental Results}
\label{sec:results}



In this section, we present detailed study of the SMSPD response to proton, electron, and pions, including results on the sensor signal properties, time resolution, and detection efficiency. 
The results from proton beams are shown in Section~\ref{ssec:results_proton} and the results from the secondary electron and pion beams are shown in Section~\ref{ssec:results_e}.

 

\subsection{Experimental Results for Protons}
\label{ssec:results_proton}
\begin{figure}[ht]
	\centering
        \includegraphics[width=0.45\linewidth]{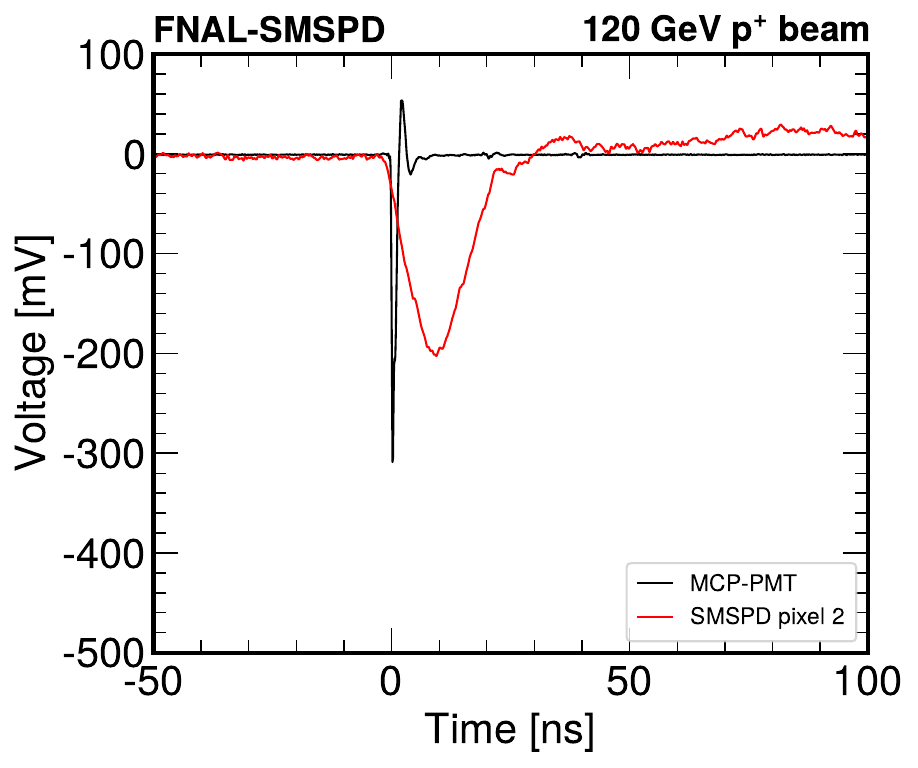}
	\includegraphics[width=0.45\linewidth]{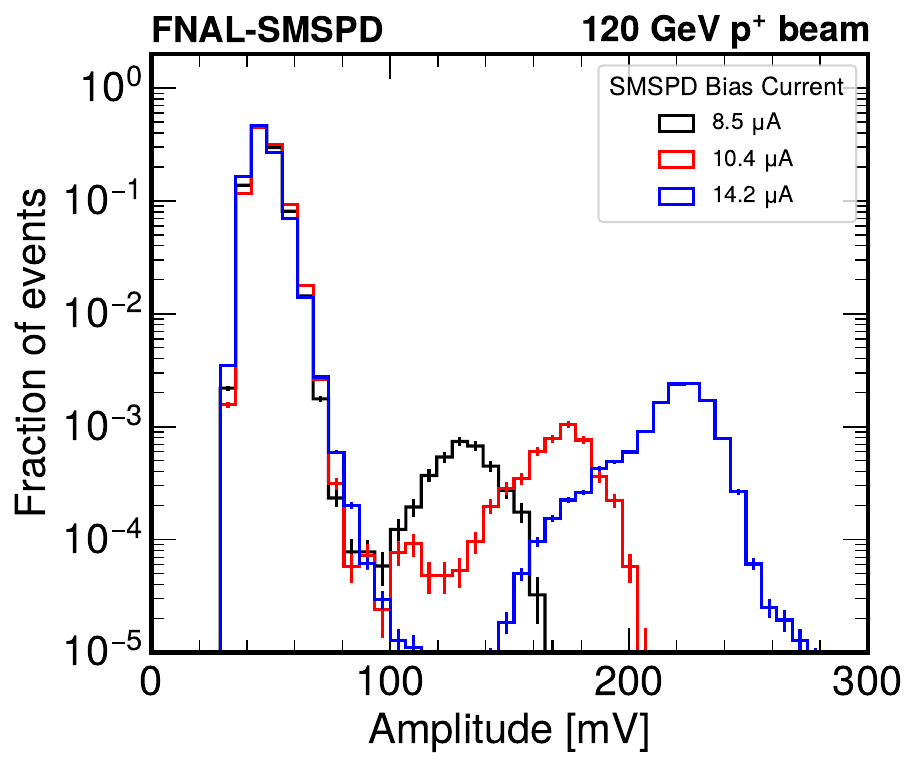}
 	\caption{The average pulse shapes (left) from the MCP-PMT and pixel 2 of the SMSPD from triggered events are shown. The reconstructed amplitude distributions (right) from the pulse shapes of SMSPD pixel 2 for various bias voltages are shown, where the peaks on the left are from baseline electronic noise and the peaks on the right are from signal pulses. 
   The mean of the signal amplitude increases linearly with the bias current as expected.
}
	\label{fig:SignalWaveforms}
\end{figure}

In this section, we will first present our result for proton beams.
Events considered in our analysis are required to have a high-quality track that has a position measurement consistent with the SMSPD area and a MCP-PMT signal consistent with a minimum ionizing particle (MIP) signal amplitude to ensure reliable references for particle position and time.
A high-quality proton track must have at least 13 hits out of the 16 strip and pixel planes of the telescope with at least four of the hits in pixel layers. 
The reduced $\chi^2$ of the track must be less than 5.
Averaged waveforms for triggered events produced by one of the SMSPD pixels in coincidence with the MCP-PMT are shown in Figure~\ref{fig:SignalWaveforms} (left). 
The waveforms for each pixel have similar shapes, while the reconstructed amplitude of the waveforms increases linearly as the bias current increases, as shown in Figure~\ref{fig:SignalWaveforms} (right). 
In order to reject signal from noise from every operating bias current, the amplitude threshold is determined for each bias current to be at least 7 standard deviation away from the noise peak.

\begin{figure}[htb!]
	\centering
	\includegraphics[width=0.49\linewidth]{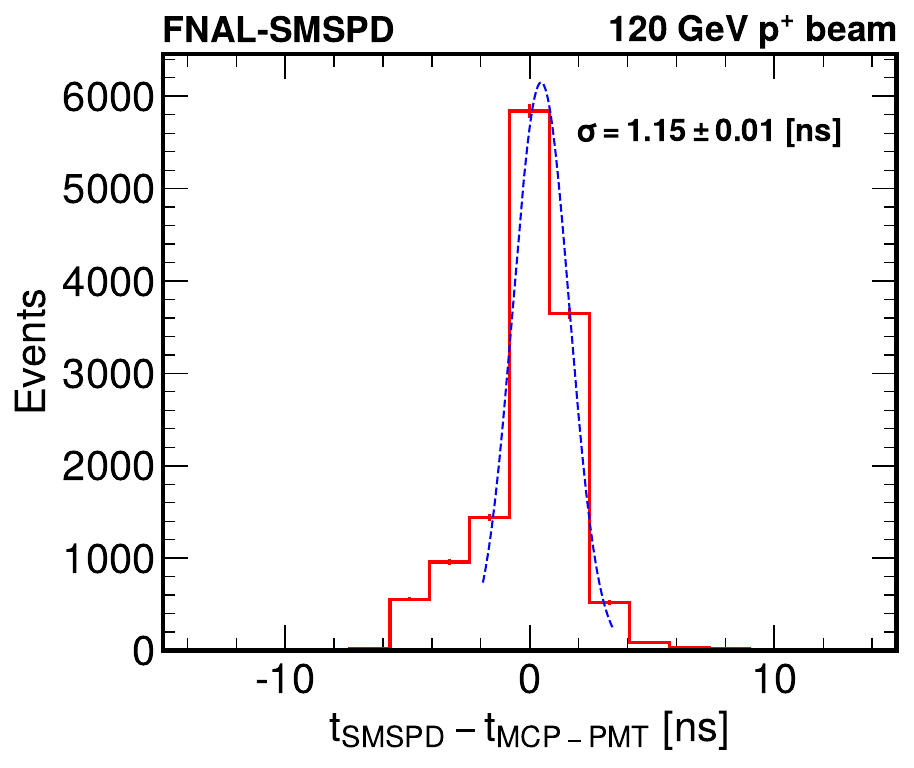}
 	\caption{The time difference between the time of arrival of SMSPD pixel 2 signal (t$_\mathrm{SMSPD}$) and MCP-PMT signal (t$_\mathrm{MCP-PMT}$) is shown, demonstrating a $1.15\pm0.01$~ns time resolution of the SMSPD.
    Pixel 2 was biased at 14.2~$\mu$A, and similar time resolution observed across different pixels and different bias current.
  }
	\label{fig:TimeResolution} 
\end{figure}

The time stamps of the SMSPD and MCP-PMT signal are both determined by performing a fit to the rising edge of the pulse to extract the time at which the pulse reaches 50\% of the maximum amplitude.
The time difference between the MCP-PMT and one of the SMSPD pixels is shown in Figure~\ref{fig:TimeResolution} and observed to be similar across different SMSPD pixels and different values of SMSPD bias current.
Since the MCP-PMT has been measured to have a time resolution of less than 10~ps~\cite{Ronzhin:2015idh}, the width of the distribution is dominated by the time resolution of the SMSPD, which is measured to be $1.15\pm0.01$~ns.
Both the SMSPD array and the readout chain were not specifically optimized for precise time resolution, so the relatively high time resolution of 1~ns is expected and consistent with previous measurements of SMSPD arrays using shorter and narrower microwires of about 325~ps with photons~\cite{10.1063/5.0150282}.
The measured time resolution includes the electronic jitter due to the readout chain and the geometric jitter caused by time delays introduced by particle detections at different longitudinal positions. The latter effect is expected to become significant as the length of the microwire increases~\cite{Korzh:2020}.
In the future, we plan to improve the time resolution by engineering devices with faster rise times and optimizing the readout scheme to correct for the longitudinal geometric jitter.

\begin{figure}[ht]
	\centering
        \includegraphics[width=0.45\linewidth]{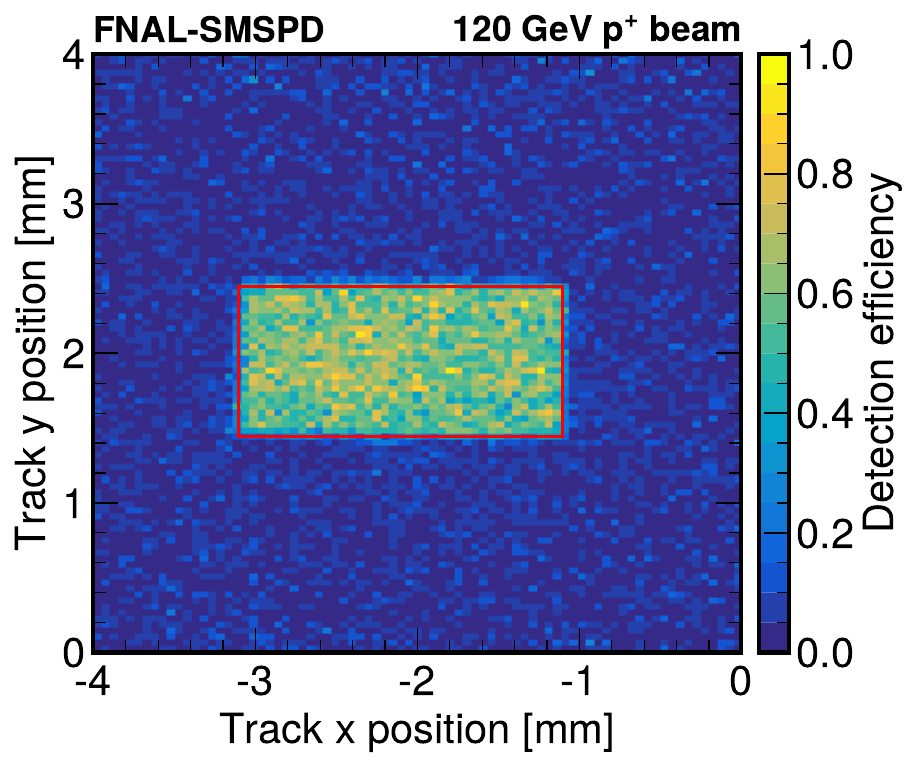}
 	\includegraphics[width=0.45\linewidth]{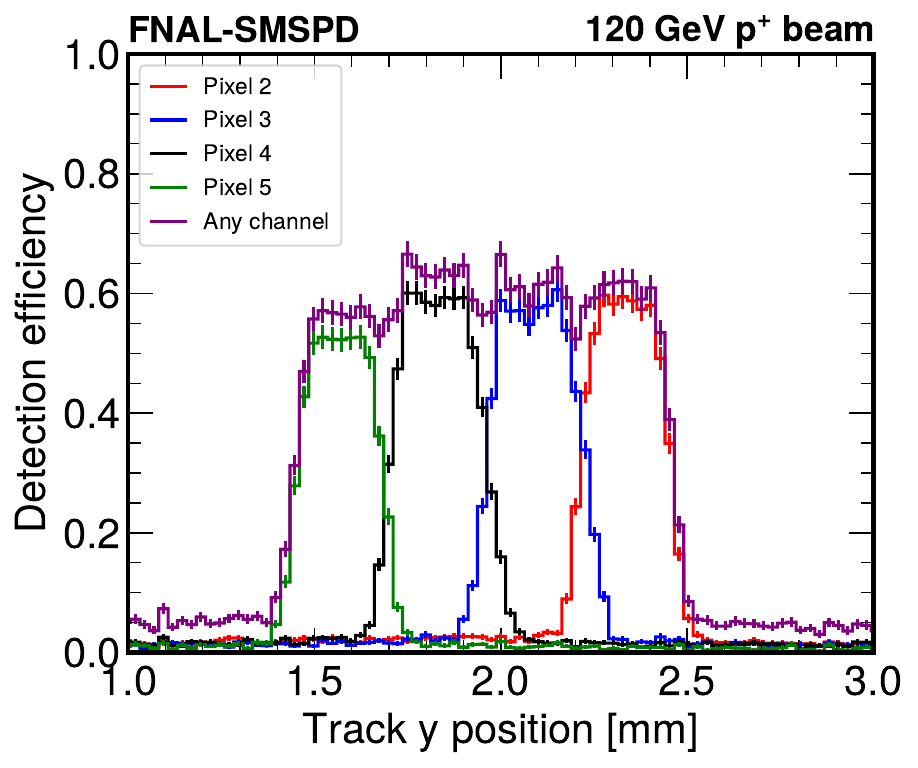}
  
 	\caption{The detection efficiency normalized to the fill factor (40\%) of the four SMSPD pixels are shown. 
  The efficiency as a function of the incident proton track position in the x and y directions (left) and in the y direction only (right). 
  The data in both plots were recorded with pixel 2 operating at 14.2~\uA, pixel 3 operating at 14.8~\uA, pixel 4 operating at 14.8~\uA, and pixel 5 operating at 14.4~\uA.
  }
	\label{fig:SNSPDEfficiency}
 
\end{figure}

In order to measure the device efficiency, events are required to have a well measured proton track in the FTBF telescope, with a position consistent with the sensor active area. The detection efficiency is measured as the fraction of these events which also has a signal above threshold in an SMSPD pixel. 
The proton detection efficiency of the SMSPD is measured with respect to the incident particle position. 
Events considered in the efficiency measurement are required to have a high-quality track with a position consistent with the sensor active area and time stamp within 5~ns from the time stamp of a MIP-like signal in the MCP-PMT. 
The detection efficiency is measured as the fraction of these events that also have a signal above threshold in one of the SMSPD pixels.
Figure~\ref{fig:SNSPDEfficiency} (left) shows the proton detection efficiency if there is a signal in any of the four SMSPD pixels as a function of the measured track x and y position. 
Figure~\ref{fig:SNSPDEfficiency} (right) shows the detection efficiency with respect to the track y position by integrating the 2-mm length of each SMSPD pixel. 
Clear contribution can be seen from each individual pixel, showing good uniformity across the pixels.

\begin{figure}[ht]
	\centering
  \includegraphics[width=0.45\linewidth]{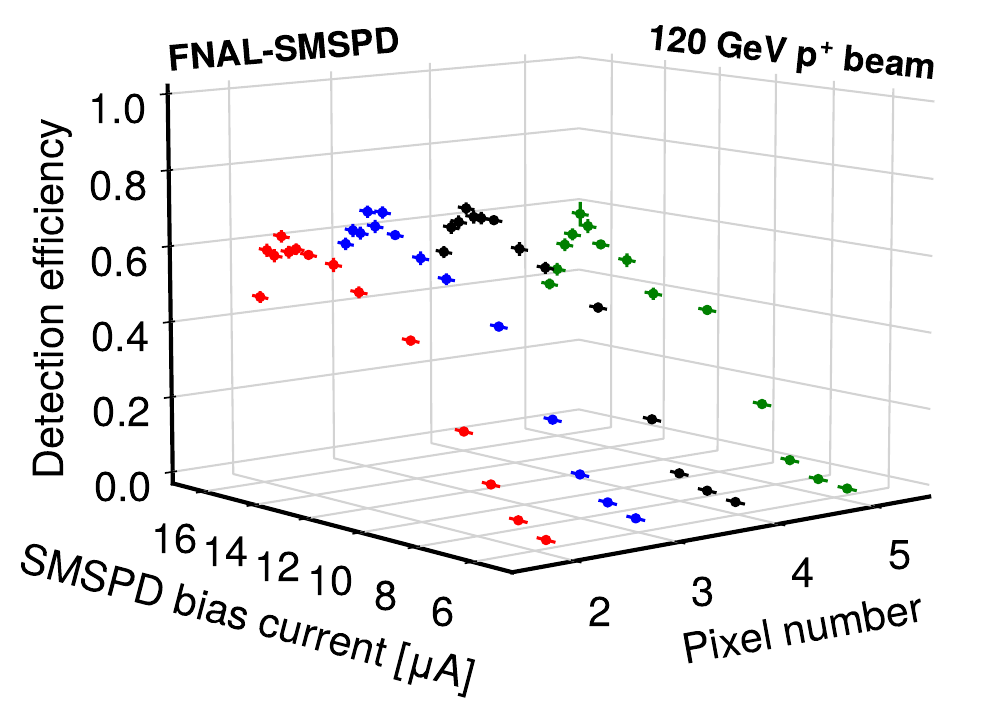}
  \includegraphics[width=0.45\linewidth]{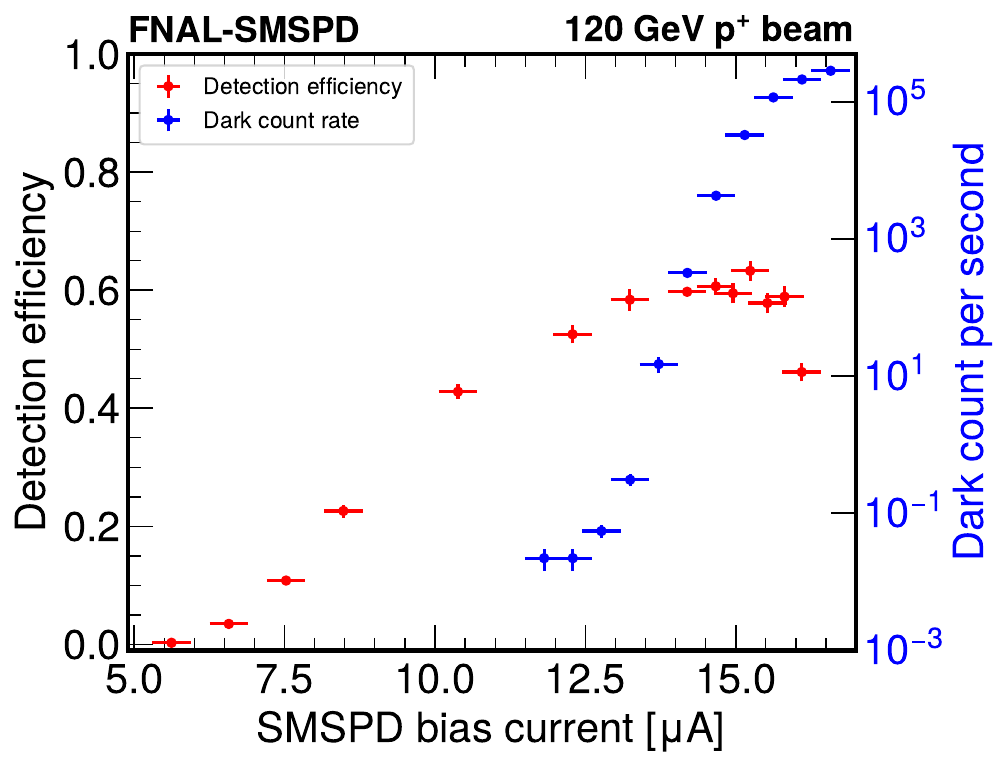}
 	\caption{The detection efficiency normalized to the fill factor (40\%) of all four measured SMSPD pixels (left) and the detection efficiency and dark count rate of pixel 2 (right) are shown.
  A correction factor of 1.117 has been applied to the inclusive detection efficiencies, due to underestimation caused by the track spatial resolution.
  The detection efficiency measurements include a relative systematic uncertainty of 0.2\% and the statistical uncertainty of each measurement.
}	\label{fig:SNSPDEfficiencyVsBias_proton}
 
\end{figure}

Finally, the inclusive detection efficiency of each of the four pixels, calculated by integrating the active area of each SMSPD pixel are shown in Figure~\ref{fig:SNSPDEfficiencyVsBias_proton} (left).
Similar and consistent efficiencies are observed across the four readout pixels.
The non-negligible 30~\um spatial resolution of tracks, which smears the turn-on of the SMSPD response at the edge of the active area, causes an underestimation of the inclusive detection efficiency when integrating the area of each pixel.
Through simulation with toy models, we have modeled the smearing and estimated that a 30~\um$\pm$ 1~\um spatial resolution would cause an underestimation of $11.7\%\pm 0.2\%$ for each pixel.
Therefore, we apply a correction factor of 1.117 to all efficiency measurements and the uncertainty on the spatial resolution of the tracks are propagated as a source of systematic uncertainty, amounting to a relative uncertainty of 0.2\% on the efficiency measurements.
An uncertainty on the SMSPD bias current, amounting to 0.32$\mu$A, is also included in Figure~\ref{fig:SNSPDEfficiencyVsBias_proton}.
The uncertainty is due to the variations of the DC offset in the DC-coupled cryogenic amplifier and the parasitic resistance of the SMSPD across different cool down cycles, as described in Section~\ref{sec:snspd}.
The detection efficiency of one of the SMSPD pixels (channel 2) is overlayed with its dark count rate, as shown in Figure~\ref{fig:SNSPDEfficiencyVsBias_proton} (right).
The dark count rate is measured with a Swabian Time Tagger about 20 hours after the beam is off.
As shown in the figure, there exists an operating regime with bias current from 10--13~$\mu$A, where the SMSPD efficiency is above 40\% and dark count rate is below 1~Hz.

\subsection{Experimental Results for Electrons and Pions}
\label{ssec:results_e}

In this section, we will present our result for the 8~GeV secondary electron and pion beams from FTBF.
To distinguish electron and pions from the secondary beam in FTBF, three radiation lengths of tungsten plates were added to the front surface of the MCP-PMT to induce showers from electrons, while pions, due to their higher masses and lower velocity, will still deposit MIP-like signals.
Therefore, a MCP-PMT signal consistent with a MIP signal amplitude is identified as pion and a MCP-PMT signal consistent with a shower signal amplitude, six times larger than the MIP signal amplitude, is identified as a showering electron.

\begin{figure}[htb]
	\centering
	\includegraphics[width=0.48\linewidth]     {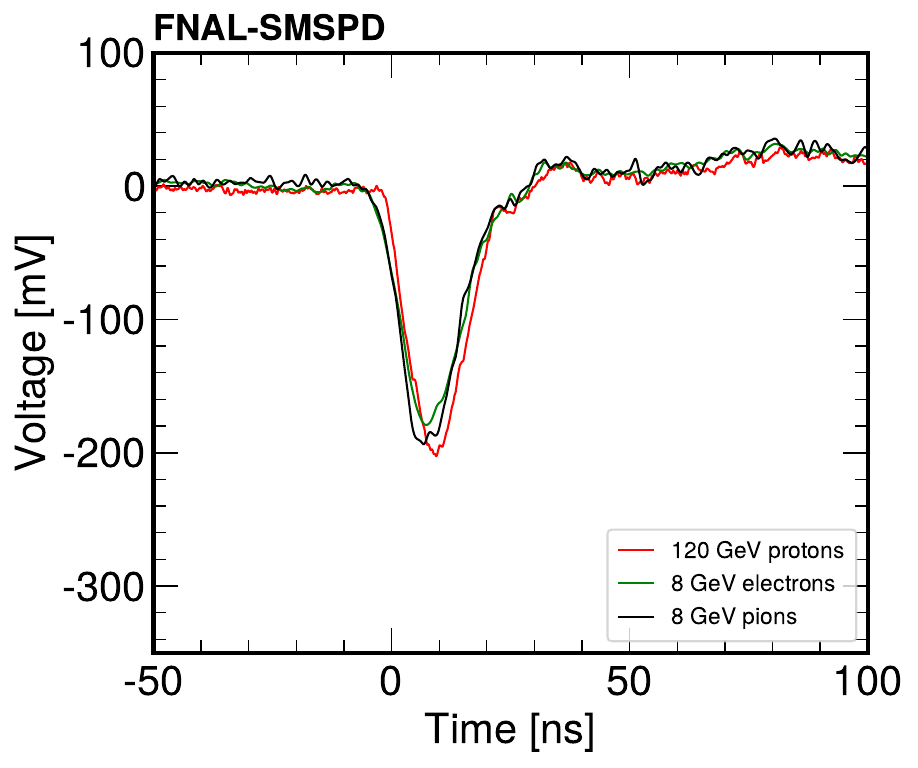}
	\includegraphics[width=0.48\linewidth]{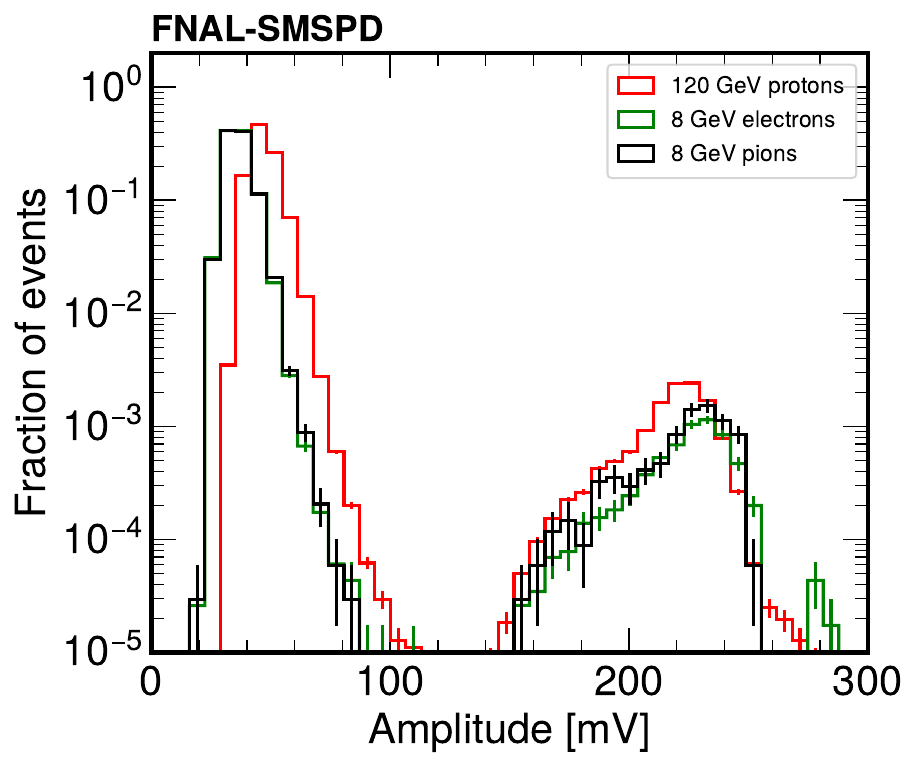}
 	\caption{The average pulse shapes (left) from the pixel 2 of the SMSPD for a triggered proton, electron and pion events are shown. The reconstructed amplitude distributions (right) from the pulse shapes of SMSPD pixel 2 for proton, electron and pion events are shown, where the peaks on the left are from baseline electronic noise and the peaks on the right are from signal pulses.
    The data in both plots were recorded with pixel 2 operating at 14.2~\uA. During the proton (electron and pion) run the analog bandwidth of the data acquisition was 2 GHz (0.35 GHz).
}
	\label{fig:SignalWaveformsElectrons}
\end{figure}

Events considered in the analysis for electrons and pions are required to have a high-quality track that has a position measurement consistent with the SMSPD active area and a MCP-PMT signal corresponding to a MIP for pions or shower for electrons.
Due to the higher scattering probability of the lower energy electrons and pions and the large amount of materials in the cryostat, only the six strip planes that are upstream from the cryostat are used to reconstruct tracks.
A high-quality track for electrons and pions must have hits in at least 5 strip planes and a reduced $\chi^2$ less than 5.
The SMSPD waveforms and amplitude distributions from electrons and pions are similar to that of protons, as shown in Figure~\ref{fig:SignalWaveformsElectrons}. 
Therefore, the same amplitude threshold is used in the data analysis of electron and pion beams.
Additionally, the time resolution of the SMSPD for 8~GeV electrons and pions are also shown to be consistent with that of 120~GeV protons.

\begin{figure}[htb!]
	\centering
    \includegraphics[width=0.60\linewidth]{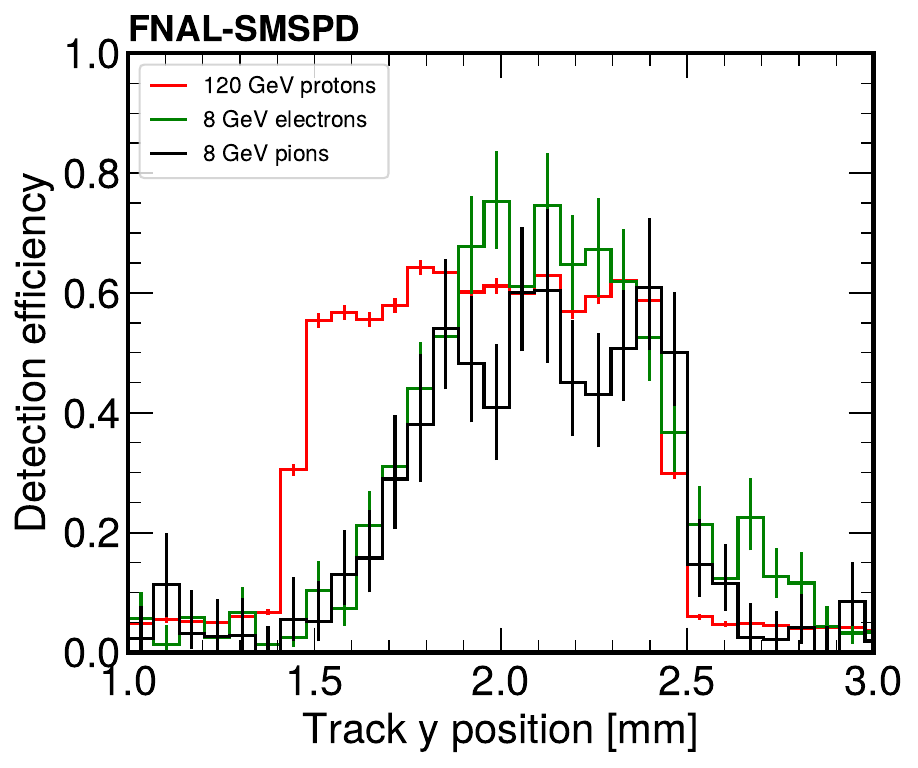} 
 	\caption{The detection efficiency normalized to the fill factor (40\%) of four SMSPD pixels as a function of the incident proton track position in the y direction for 120~GeV protons (same as what is shown in the right panel of Figure~\ref{fig:SNSPDEfficiency}), 8~GeV electrons, and 8~GeV pions are shown.
    The bottom-most pixel with the lowest y position (pixel 5) was not biased for electron and pion beams, resulting in a smaller active area in the y-direction.
    The data were recorded with pixel 2 operating at 14.2~\uA, pixel 3 operating at 14.8~\uA, pixel 4 operating at 14.8~\uA, and pixel 5 operating at 14.4~\uA (only for proton beam).
    The error bars on the data points only include the statistical uncertainty.}
	\label{fig:SNSPDElectronEfficiency}
 
\end{figure}

The detection efficiencies of electron and pion along with proton with respect to the track position in y are shown in Figure~\ref{fig:SNSPDElectronEfficiency}. 
Only three out of the four pixels that were read out for proton beams were read out for electron and pion beams, due to a malfunction of the power supply biasing pixel 5 (bottom-most pixel) during the test beam campaign, resulting in a smaller active area in the y-direction.
Additionally, lower energy pions and electrons are more likely to scatter in the tracker, resulting in lower statistics and a worse track resolution of 130~\um measured from fitting the turn-on of the SMSPD response in-situ.
Therefore, the edge of the SMSPD is more smeared for electrons and pions.
However, it can be seen for the first time that the SMSPD detection efficiency is similar across the three particle types.

\begin{figure}[htb!]
	\centering
	\includegraphics[width=0.7\linewidth]{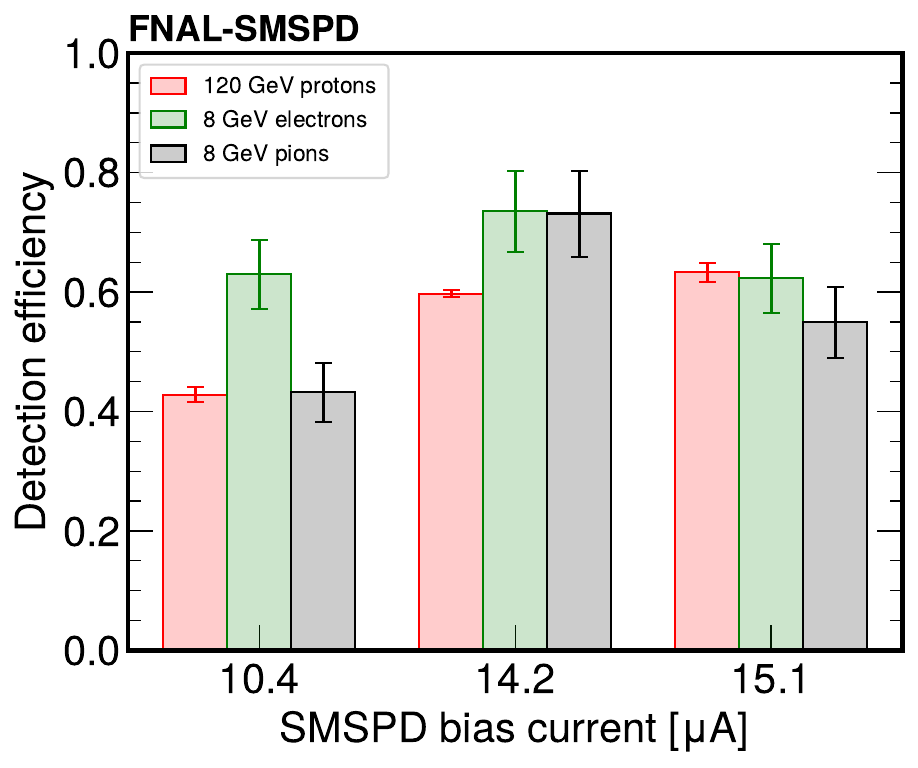}
 	\caption{The detection efficiency for proton, electron, and pion with respect to the bias current.
  A correction factor of 1.8 (1.117) has been applied to the inclusive detection efficiencies for electrons and pions (protons), due to underestimation caused by the track spatial resolution.
  The detection efficiency measurements include a relative systematic uncertainty of 7.8\% (0.2\%) for electrons and pions (protons) and the statistical uncertainty of each measurement.
  }
  \label{fig:SNSPDEfficiencyVsBias}
\end{figure}

Finally, the inclusive detection efficiency per pixel for electron and pion, calculated by integrating the active area of each SMSPD pixel, is compared against the proton detection efficiency at three different operating bias currents, as shown in Figure~\ref{fig:SNSPDEfficiencyVsBias}.
Using the same simulation model as mentioned in the previous section, the 130~\um$\pm$20~\um spatial resolution of the tracks would result in an 80\%$\pm$14\% underestimation in the inclusive detection efficiency measurement.
Therefore, a correction factor of 1.8 and a relative systematic uncertainty of 7.8\% is applied to the electron and pion efficiencies in Figure~\ref{fig:SNSPDEfficiencyVsBias}.
It can be seen from the plot that for the first time, we observed that there is no significant difference of the SMSPD detection efficiency across the three particle types at all three measured bias currents.

\section{Summary}
\label{sec:summary}
In summary, this paper presents detailed studies of an 8-channel $2\times2$~mm$^{2}$ WSi SMSPD array exposed to 120~GeV proton beam and 8~GeV secondary electron and pion beam at the FTBF.
The absolute and calibrated detection efficiency was measured for the first time for protons, electrons, and pions, enabled by the strip and pixel tracker that provided precise spatial resolution of 30~\um for 120~GeV protons and 130~\um for 8~GeV electrons and pions.
The result demonstrated consistent detection efficiency across pixels and at different bias currents.
Time resolution of 1.15~ns was measured for the first time for SMSPD with proton, electron, and pions, enabled by the use of MCP-PMT which provided ps-level reference time stamp.
The results presented in this paper is the first step towards developing SMSPD array systems for particle detection and identification at the next-generation accelerator-based experiments.

In the future, we plan to develop superconducting wire detectors that are more optimized for high energy particle detection with ultra-precise spatial and temporal resolution and large active areas.
Additionally, we plan to improve our characterization system in FTBF by increasing the number of readout channels and improving the mechanical structure to place the telescopes more optimally with respect to the telescope to achieve the design spatial resolution of 8~\um per track~\cite{KWAN2016162}.
Finally, the following test beam campaigns will aim to fully characterize superconducting wire detectors with the other secondary beams available at FTBF, including electrons, pions, and muons with energy ranging from 1--60~GeV, and in a high-rate and high-occupancy environment at the Fermilab Irradiation Test Area for the first time.
These upcoming test beam campaigns will enable us to study the feasibility of using superconducting wire detector arrays in future collider or beam-target experiments.

\acknowledgments

This manuscript has been authored by Fermi Research Alliance, LLC under Contract No. DE-AC02-07CH11359 with the U.S. Department of Energy, Office of Science, Office of High Energy Physics.
C.W., S.X., A.B., M.S. are partially supported by the U.S. Department of Energy, Office of Science, Office of High Energy Physics, under Award No. DE-SC0011925.
C.P., S.X., A.A. are partially supported by the U.S. Department of Energy, Office of Science Accelerate Initiative Program Award under FWP FNAL 23-30. This work was partially supported by the Fermilab New Initiatives Program. 
V.V., M.B., and C.S. are supported by the Chilean ANID PIA/APOYO AFB230003.
V.V. is partially supported by the ``Programa de Incentivo a la Investigaci\'on Cient\'ifica'' (PIIC) of the UTFSM.
Part of this research was performed at the Jet Propulsion Laboratory, California Institute of Technology, under contract with the National Aeronautics and Space Administration.

We thank the Fermilab accelerator and FTBF personnel for the excellent performance of the accelerator and support of the test beam facility, in particular M. Kiburg, E. Niner, N. Pastika, T. Nebel, and E. Schmidt. We also thank L. Uplegger for developing the telescope tracker system and providing late night supports. We thank C. Madrid for his technical support during the operation and data taking time.

\bibliography{refs}
\bibliographystyle{jhep}
\end{document}